\documentclass[aps, prl, reprint, longbibliography]{revtex4-1}
\usepackage{graphicx,amsmath}
\usepackage{subfig}
\usepackage{color}
\DeclareGraphicsExtensions{.pdf,.eps,.png,.jpg,.mps}
\graphicspath{ {Figures/} }

\begin{document}

\title{Analog control with two Artificial Axons}
\author{Hector G. Vasquez}
\author{Giovanni Zocchi}
\email{zocchi@physics.ucla.edu}
\affiliation{Department of Physics and Astronomy, University of California - Los Angeles}

\begin{abstract}
\noindent The artificial axon is a recently introduced synthetic assembly of supported lipid bilayers 
and voltage gated ion channels, displaying the basic electrophysiology of nerve cells. Here we 
demonstrate the use of two artificial axons as control elements to achieve a simple task. 
Namely, we steer a remote control car towards a light source, using the sensory input dependent 
firing rate of the axons as the control signal for turning left or right. We present the result in the form 
of the analysis of a movie of the car approaching the light source. In general terms, 
with this work we pursue a constructivist approach to exploring the nexus 
between machine language at the nerve cell level and behavior.  
 
\end{abstract}

\maketitle

\noindent {\bf Introduction.} 
Action potentials form the machine language of the nervous system. While the mechanisms 
and characteristics of action potentials have been known for over half a century, 
these most interesting excitations have not been produced in vitro, in a synthetic biology 
setting, until very recently \cite{Hector2017}. The artificial axon (AA) 
\cite{Amila2016, Hector2017} is an excitable node which supports action potentials, 
that is, voltage spikes produced by the same ionic mechanism as action potentials 
in neurons. Our aim is to develop it into a platform for synthetic biology constructions 
which echo neuronal systems. Since the AA 
works, as far as electrical excitability, through the same physical mechanism as the real 
neuron, it can be endowed with some of the same capabilities, and also 
suffers from similar constraints. Learning to build useful networks within these constraints 
is a viable engineering approach to understanding 
brain function. Existing constructive approaches are computational \cite{IBM2009}, 
electronic \cite{Indiveri2011, Pfeil2013}, or based on real neurons, 
directing their pattern of connection \cite{Wyart2002, Elisha2005, Elisha2006, Chang2006, Elisha2008, 
Genovesi2012}, and re-programming stem cells in 3D cultures \cite{Pasca2015}. 
Especially the first two are much more advanced than what  
we show here; they also have a longer history, and, crucially, they are based 
on electronics. Specifically, neuromorphic chips (NMCs) are currently competitive with traditional 
von Neumann architecture neural networks in solving complex problems in pattern recognition \cite{Schuman2017}. 
However, NMCs are {\it simulators} of spiking nerves, based on a fundamentally 
different microscopic process, namely electronics. Although the higher level architecture 
may be designed similar to a real network of neurons, the microscopics is different. 
Cultured neurons have been used by the Moses group to construct logic functions such as AND, 
also demonstrating a remarkable reliability achieved through a redundancy of connections \cite{Elisha2008}. 
Our approach is an attempt to simplify the patterned neuron paradigm, by introducing 
a simpler, synthetic ``neuron". In contrast to NMCs, AAs are to be based on the same 
microscopic process as real neurons, namely ionics. 
This is, we believe, an incisive experimental approach to study how the microscopics of neurons  
may generate complex macroscopic responses, patterns, and behaviors. The latter program 
partakes of the underlying thread of condensed matter physics, and our approach is informed 
by that discipline. \\ 
Coming back to NMCs, interest in the field is both scientific and practical, 
the latter because NMCs based AI may enjoy possibly orders of magnitude better energy 
consumption characteristics compared to von Neumann architecture AI. 
AA chips could in principle enjoy similar architectural advantages, plus additional power 
saving benefits due to the fact that ionics circuitry works at $\sim 100 \, mV$ 
vs the $\sim 1 \, V$ of electronic circuitry \cite{Hector2017}. However, it should be mentioned 
that where power density is the sole criterion, notwithstanding, for example, portability, 
there are low temperature devices which can fare much better \cite{Schneider2018}. 
Our objective with the AA in the near future is to develop a breadboard or ``tool kit" with which to construct task performing networks which are based on the same microscopics as neurons, 
and therefore are ``possible brains". At the moment, even one AA is sufficiently delicate,  
and complicated to make, that we thought it would be useful to give a demonstration of an 
actual task performing device based on AAs, even though we can only make 
very simple ones at present. Namely, we use a system of two AAs to steer a remote control 
toy car towards a light source.  

\begin{figure}
\centering
\includegraphics[width=3in]{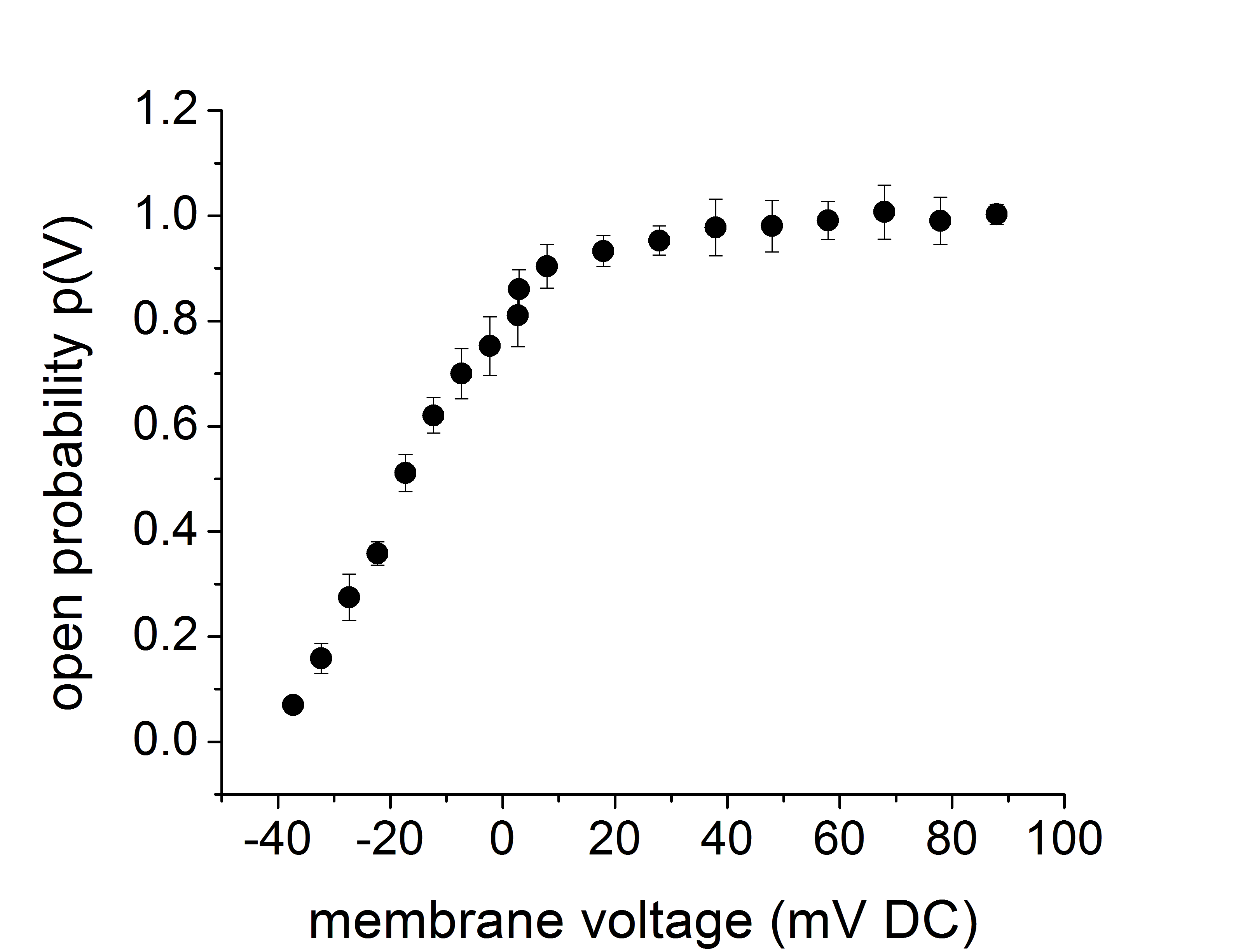}
\captionsetup{justification=raggedright, singlelinecheck=false}
\caption{Opening probability for the potassium channel used in this study (KvAP); 
reproduced from \cite{Amila2013}.}
\label{fig:openp_KvAP}
\end{figure}

\noindent First, we briefly describe the AA, which was introduced previously \cite{Amila2016, Hector2017}, 
and give an overview of the present system. In the next section, we describe the results: they consist of 
a demo in the form of a movie, with accompanying analysis. In Materials and Methods we describe in detail 
the EE aspects, in order to report exactly what we do with the AAs and the electronic circuitry. 
We conclude with a brief discussion of relevance and scope. \\ 
The present AA is an excitable $100 \, \mu m$ size ``node" consisting of a supported 
phospholipid bilayer with $\sim 100 $ voltage gated potassium channels (KvAP) embedded 
and oriented. A concentration gradient of potassium ions is maintained across the bilayer 
by means of reservoires; typically $[K^+]_{in} \approx 30 \, mM$ for the ``inside" of the 
axon and $[K^+]_{out} \approx 150 \, mM$ for the ``outside". In the following, all potentials 
represent the potantial difference between the inside and outside of the axon: 
$V = V_{in} - V_{out}$. The aforementioned ionic gradient results in an equilibrium 
(Nernst) potential 

\begin{equation}
V_N \, = \, \frac{k T}{|e|} {\text {ln}} \frac{[K^+]_{out}}{[K^+]_{in}} \, \approx \, + \, 40 \, mV
\label{eq:Nernst}
\end{equation}

\begin{figure}
\centering
\includegraphics[width=3in]{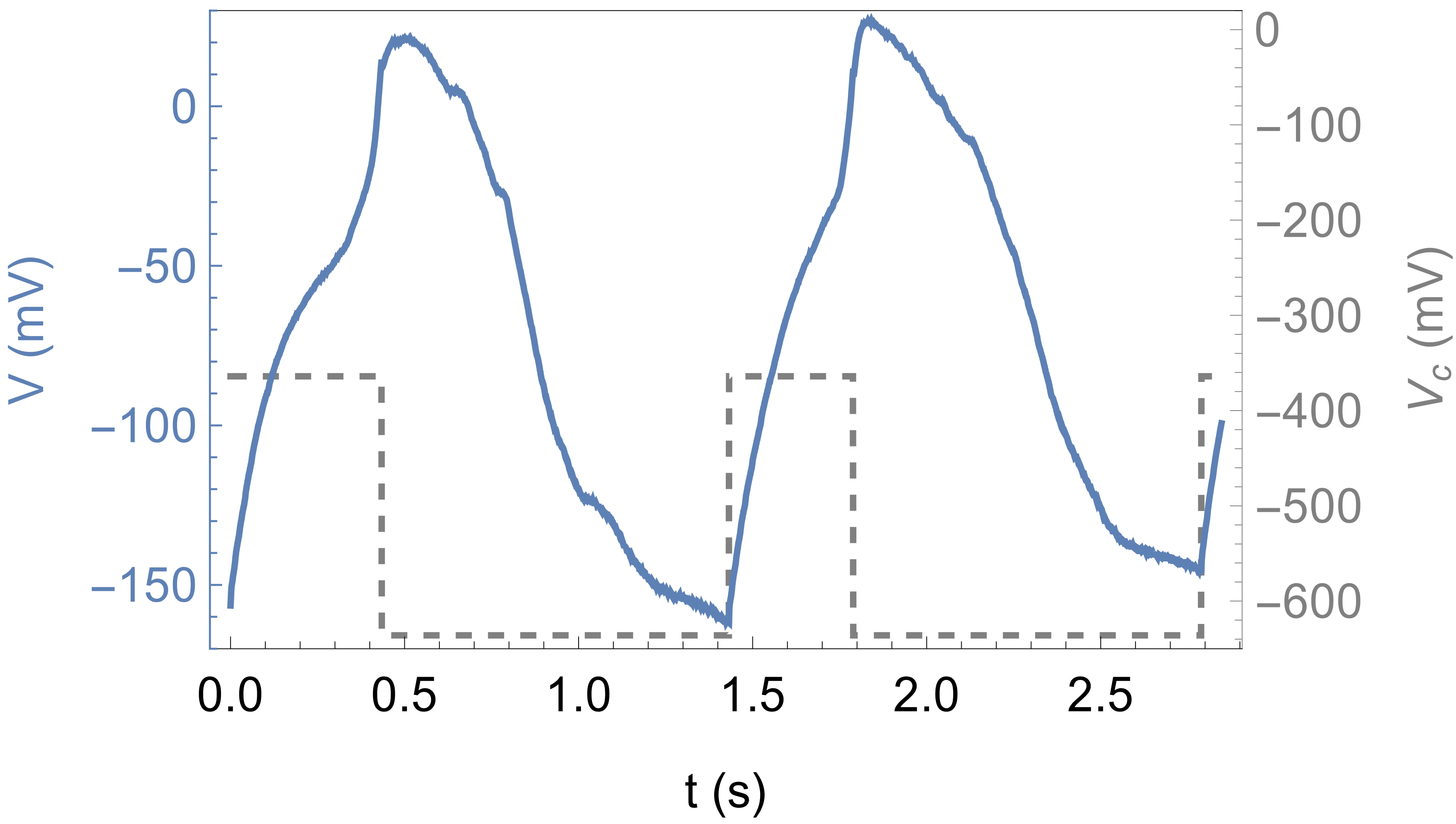}
\captionsetup{justification=raggedright, singlelinecheck=false}
\caption{Action potentials in the AA, elicited by a constant current input. Plotted is the measured 
membrane potential ($V_m$: left scale). The dotted trace shows the protocol of the CLVC 
($V_c$: right scale). The downward step in $V_c$ is triggered by $V_m$ crossing $10 \, mV$; see text for 
explanations.}
\label{fig:action_pot}
\end{figure}

\noindent Here, $k T \approx 25 \, meV$ is the thermal energy at room temperature and $e$ the 
electronic charge. In the real neuron, there is in addition an opposite gradient of sodium 
ions across the membrane, which results in a ``resting potential" $V_r$ intermediate 
between the Nernst potentials of the $K^+$ and $Na^+$ ions, and corresponding 
voltage gated sodium channels. In the AA, that role is played by the current limited voltage 
clamp (CLVC) \cite{Amila2016}, which keeps the AA voltage out of equilibrium, typically 
at a ``resting potential" $V_r \approx - 120 \, mV$. The response curve of the KvAP channels 
(Fig. \ref{fig:openp_KvAP}) shows that they are closed at $V = V_r = - 120 \, mV$, but 
if a stimulus brings the voltage above $V \sim - 20 \, mV$ the channels open, the 
chemical potential driven channel current overwhelms the CLVC current, and the AA ``fires". 
The subsequent inactivation of the KvAP channels, which stochastically enter a third, 
closed, state, allows the CLVC to pull the membrane potential back to the resting 
value. In Fig. \ref{fig:action_pot} we show a train of two action potentials in an artificial axon. 
The input stimulus is a constant current ($\sim 128 \, pA$), delivered with a current clamp. 
The membrane potential $V_m$ is measured through a separate electrode, and corresponds to the 
ordinate scale on the left in the figure. The dotted trace represents the command voltage $V_c$ to the 
CLVC, and corresponds to the ordinate scale on the right. Instead of simply keeping $V_c$ constant 
at $\sim - 364 \, mV$, which would maintain, in this case, a resting potential $V_m \approx - 120 \, mV$, 
the protocol for $V_c$ is that when the axon ``fires", $V_c$ is lowered to $- 636 \, mV$ for a fixed time 
($= 1 \, s$), then returns to $- 364 \, mV$. This maneuver is necessary but incidental to the particular 
inactivation dynamics of the ion channel we are using (the KvAP). Namely, after a firing we must pull 
the membrane potential $V_m$ down to about $- 150 \, mV$ in order that the channels re-activate 
fast enough to be ready for the next firing.  

\begin{figure}
\centering
\includegraphics[width=3in]{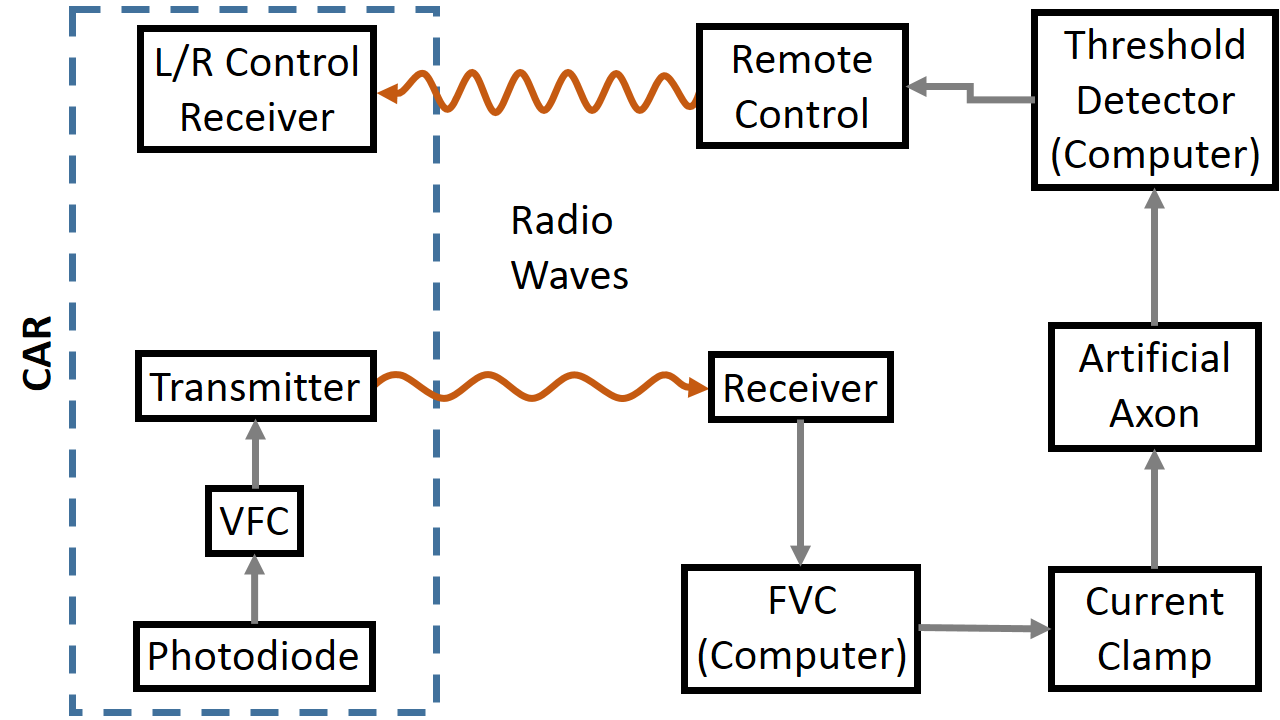}
\captionsetup{justification=raggedright, singlelinecheck=false}
\caption{System block diagram. The complete system consists of two such circuits, one 
for the right (R) photodiodes and AA and one for the left (L) components. 
VFC: voltage to frequency converter; FVC: frequency to voltage converter.}
\label{fig:block_diagram}
\end{figure}

\noindent We have previously documented some basic electrophysiology with this system, such as 
integrate-and-fire dynamics \cite{Hector2017}. For our present purpose, we use two AAs 
as a control module to steer a toy car towards a light source. Fig. \ref{fig:block_diagram} shows a diagram 
of the system. The AAs sit on an optical table in the lab; communication with the 
car is by radio waves. The remote control car is modified with two sets of photodiodes ("eyes")
accepting light from the right (R) and left (L) side of the car, respectively, and corresponding 
voltage-to-frequency converters (VFCs) and transmitters. Following for example the signal 
from the R photodiode, its voltage output is converted to frequency (in the kHz range) 
by the VFC and transmitted; this signal is received by a receiver on the optical table, 
converted to voltage by another VFC, and used as the input to a current clamp, or 
"synapse", which injects a proportional current (in the tens of pA range) into the right 
(R) axon. Action potentials in the AA trigger a threshold detector which inputs into the remote 
control module of the car the signal to turn the wheels to the right. We use the actual 
remote control of the toy car, and so the same receiver and right/left control built 
into the car. A similar but independent pathway conditions the signal from the left 
photodiode set. In summary, this system realizes a very simple analogue control protocol: 
each time the R axon spikes, the wheels of the car are turned to the right and stay there 
until the next signal comes in, and similarly for the L axon, which turns the wheels to the left. 
Action potentials in the R/L axon are induced by the light intensity falling onto the R/L 
photodiode set. So while all the peripheral systems are, at the moment, electronic, 
the "decision making algorithm" is implemented by ionics. 

There is no interaction between the two 
AAs in this realization, and the only property of the AA which we really exploit is the 
``integrate-and-fire" dynamics. The system has a degree of stochasticity (due to the 
relatively small number of ion channels in the axons), lots of noise which is not only 
thermal in origin, makes many mistakes, has many defects, and ends up looking 
``biological" (see movie). \\ 

\begin{figure}
\centering
\includegraphics[width=3in]{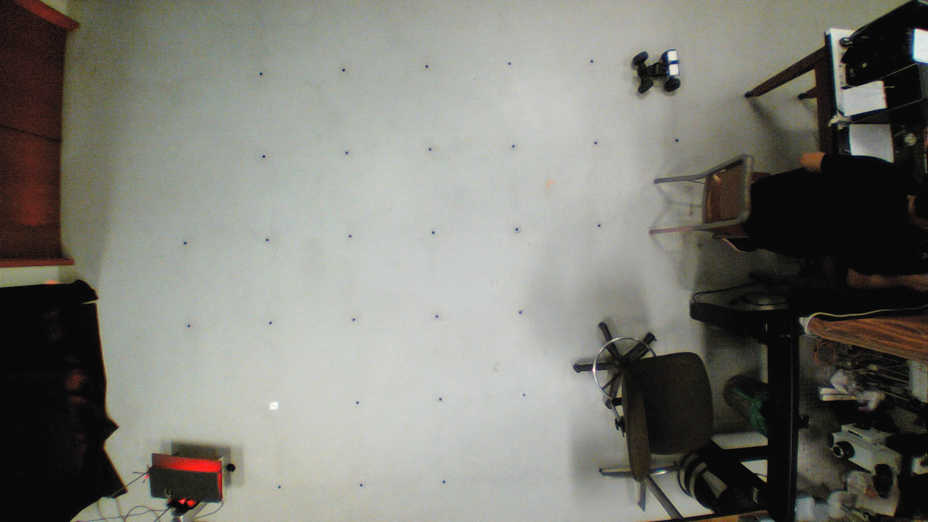}
\captionsetup{justification=raggedright, singlelinecheck=false}
\caption{Still of the room of the room of the demo, seen from the ceiling. The car and light source 
are at diametrically opposite corners. Visible on the right is the optical table with the artificial axons 
and the electronics, as well as H.G.V.}
\label{fig:the_room}
\end{figure}

\noindent {\bf Results.}  
The main result we present is a demonstration of the system in the form of the 
accompanying movie (see Ancillary files), which we now describe. 
The car moves in a (previously decluttered) laboratory room of about $5 \times 5 \, m^2$; 
in the movie, the light source is in the SW corner of the screen, and the car starts at the 
NE corner, facing W. The other bright spots on the screen are reflections of the light 
source from objects at the perifery of the room. Fig. \ref{fig:the_room} is a picture of the room seen 
from the ceiling. 

\begin{figure}
\centering
\includegraphics[width=3in]{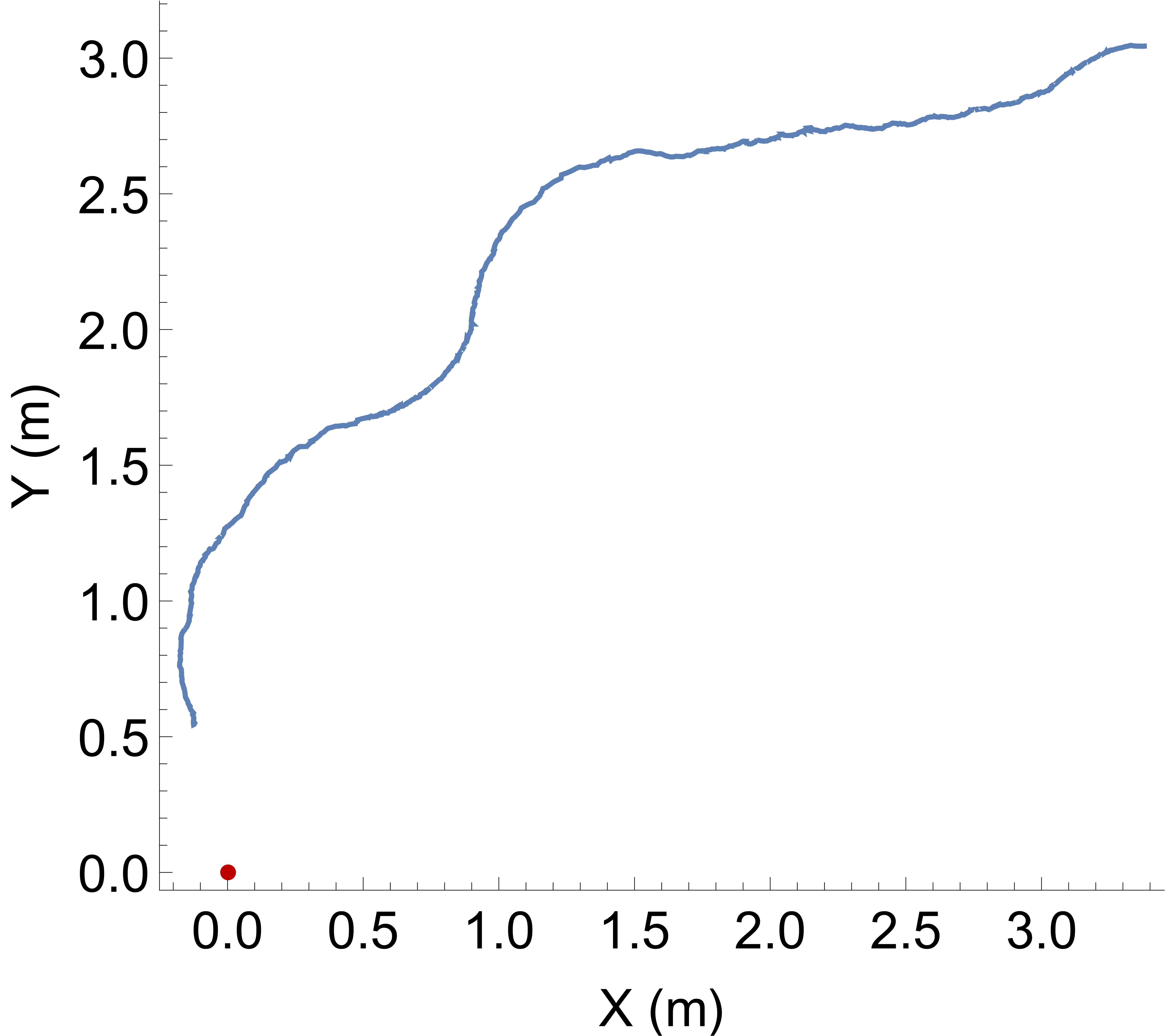}
\captionsetup{justification=raggedright, singlelinecheck=false}
\caption{Car trajectory corresponding to the run shown in the movie. 
The light source is at the origin; the scales on the axes are in $m$, and the car starts at $(x, y) = (3.4, 3.0).$}
\label{fig:trajectory}
\end{figure}

In the movie, watch the front wheels of the car 
repeatedly switch between L and R, and the overall progress. We had to slow down the 
speed of the car to match it to our rather slow axons, so we adopted short, regular spurts 
of forward motion. The actual average speed of the car is however not constant, 
as you see in the movie, because at times the tires slip on the polished floor. 
From an engineering standpoint, we may view this circumstance as simply one of many 
``defects" or sources of noise in the system. There are many such sources of randomness, 
from the microscopic scale of the individual ion channels in the AA to the macroscopic scale of 
the tires. As a result, the trajectory of the car, the ``behavior", is not deterministic 
(starting from identical initial conditions, different realizations of the car's trajectory will be 
different); however, the car does find the light source in the end. Fig. \ref{fig:trajectory} shows 
the trajectory corresponding to the movie. \\ 

\begin{figure}
\centering
\includegraphics[width=3in]{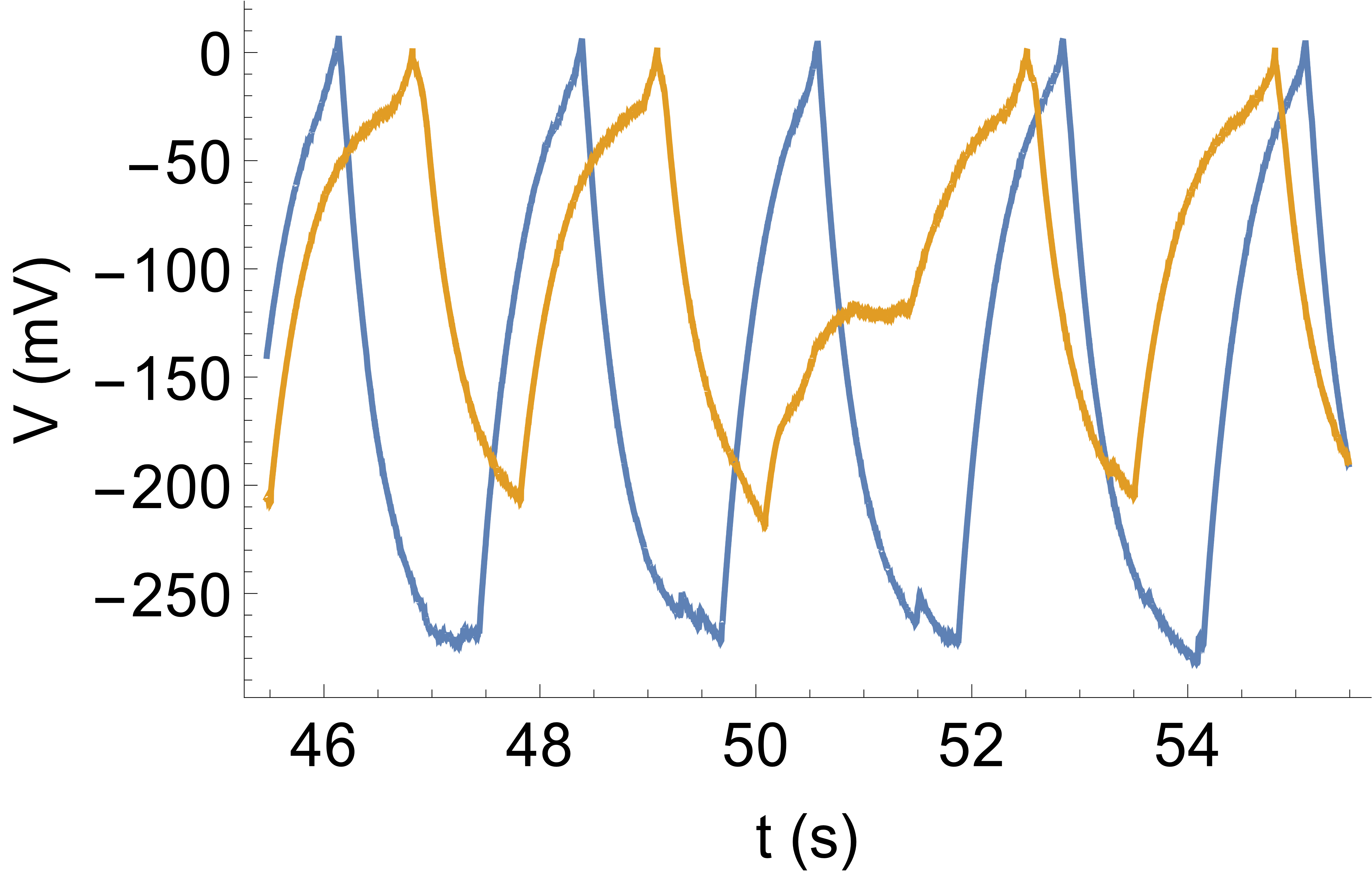}
\includegraphics[width=3in]{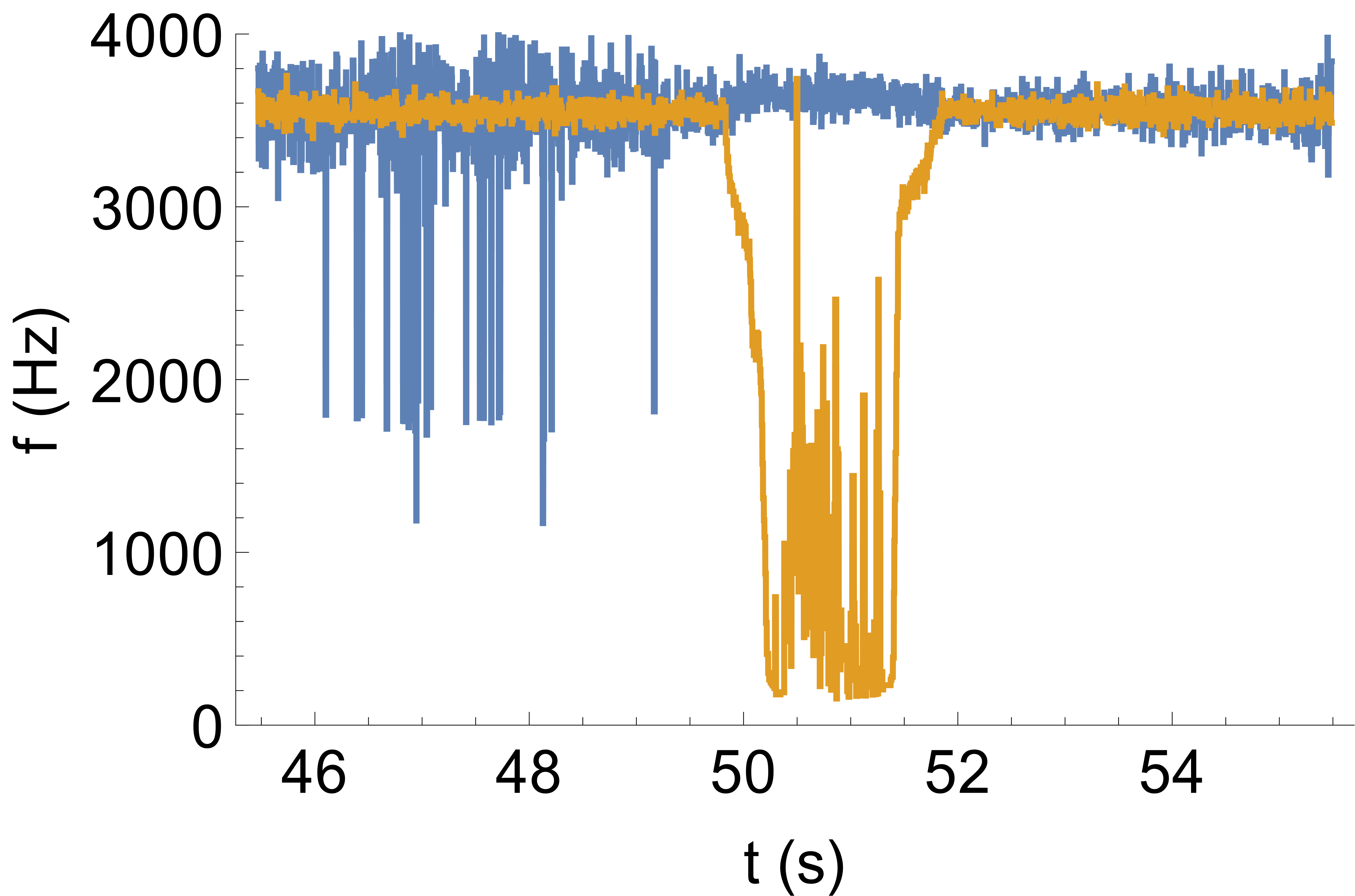}
\captionsetup{justification=raggedright, singlelinecheck=false}
\caption{(a) Membrane potential in the Left artificial axon (blue trace) and the Right AA (yellow trace) 
for part of the run shown in the movie. \\
(b) The signal at the output of the voltage to frequency converter (VFC), for the R and L circuits, and 
the same time interval as in (a). The currents injected by the synapses into the respective AAs are 
proportional to these signals.}
\label{fig:spikes_20s}
\end{figure}

Let us now describe the very simple ``machine language" with which the system operates. 
Fig. \ref{fig:spikes_20s}a shows action potentials in the two AAs over a time of 10 s; the blue trace is 
the membrane potential of the left AA, the yellow trace is the right AA. The response 
of the car is that when the blue trace crosses $4.5 \, mV$ from below, the wheels turn left  
and stay there until further notice; similarly, when the yellow trace crosses $0 \, mV$,  
the wheels turn right and stay there. What decides, then, whether overall the car 
is turning L or R is the relative phase of the spikes in the two AAs. In the example 
shown, for $45 < t < 50.5 \, s$ the car is, overall, turning R because the time interval between 
a yellow and the next blue zero crossing is larger than the time between a blue and the 
next yellow. On the other hand, for $50.6 < t < 56 \, s$ the car is overall turning L; this is 
a consequence of the R photodiodes seeing less light for $50 < t < 52 \, s$ 
(Fig. \ref{fig:spikes_20s}b), which causes a delay in the yellow spikes, changing the phase relation 
between blue and yellow spikes.
Even with identical stimuli (input currents from the ``synapses"), the firing rates of the two 
AAs are not the same (due to physical differences between the AAs, for instance, 
different number of ion channels, different leak currents, etc.). This circumstance introduces 
``phase noise", yet another source of (non-thermal) stochasticity which however does not 
prevent the overall working of the system. That is, the two AAs do not need to be perfectly 
tuned as far as firing rates. 

\begin{figure}
\centering
\includegraphics[width=3in]{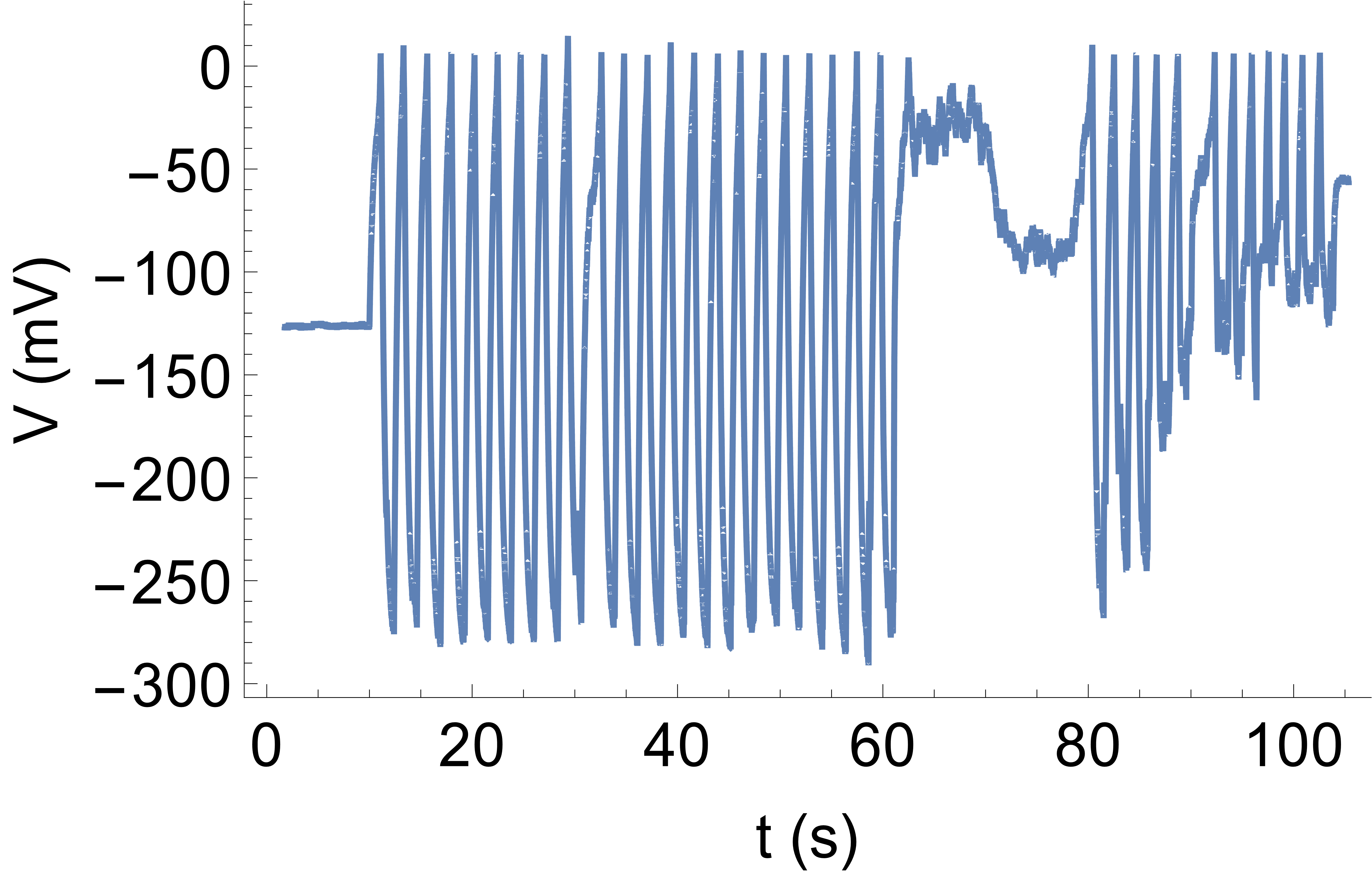}
\includegraphics[width=3in]{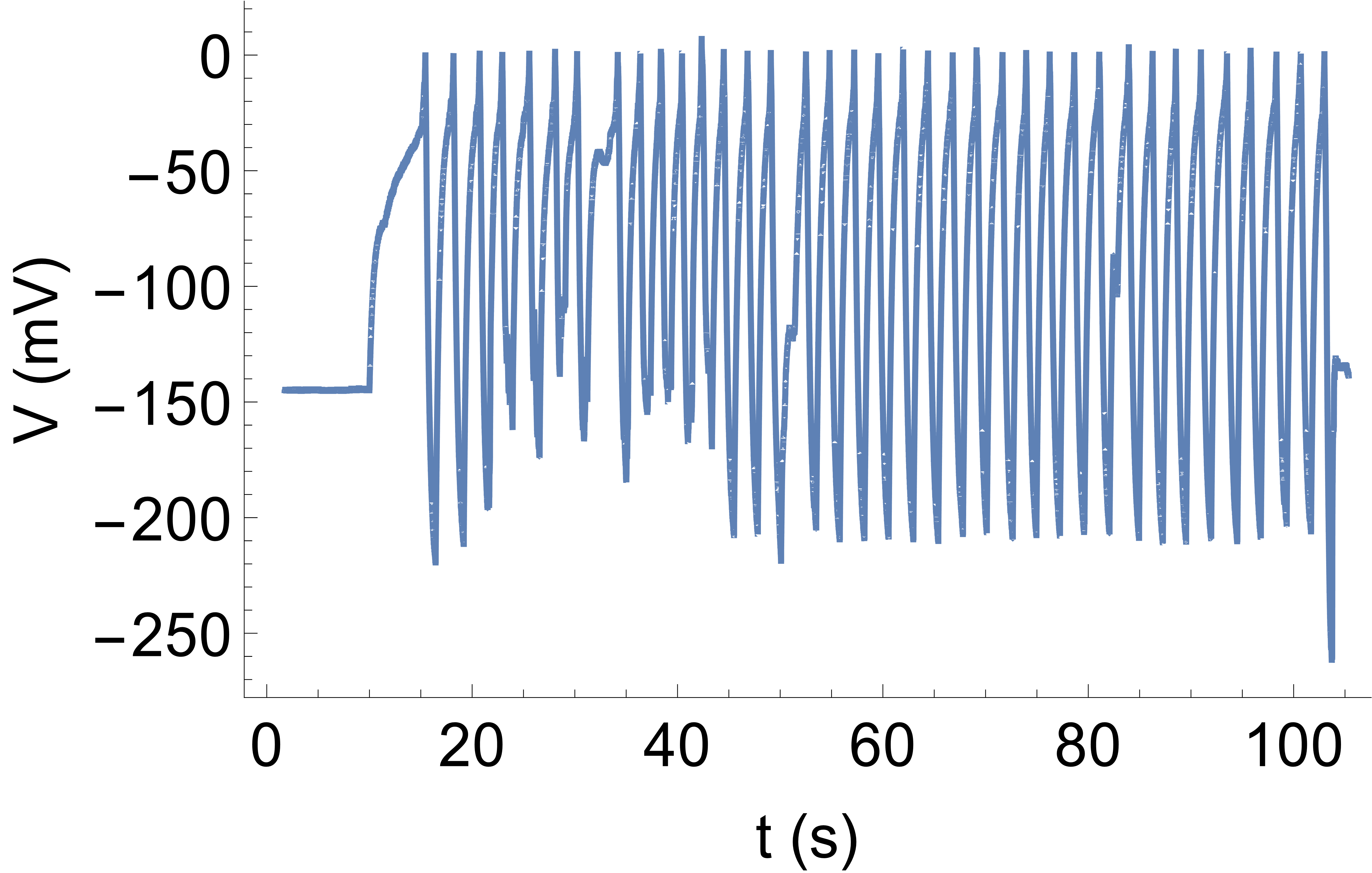}
\captionsetup{justification=raggedright, singlelinecheck=false}
\caption{Whole time series of action potentials corresponding to the run of the movie. Time $t = 0 \,s$ corresponds to the start of the video. The data recording begins 1.75 seconds later. \\
(a) Left axon; (b) right axon.}
\label{fig:spikes_all}
\end{figure}

\begin{figure}
\centering
\includegraphics[width=3in]{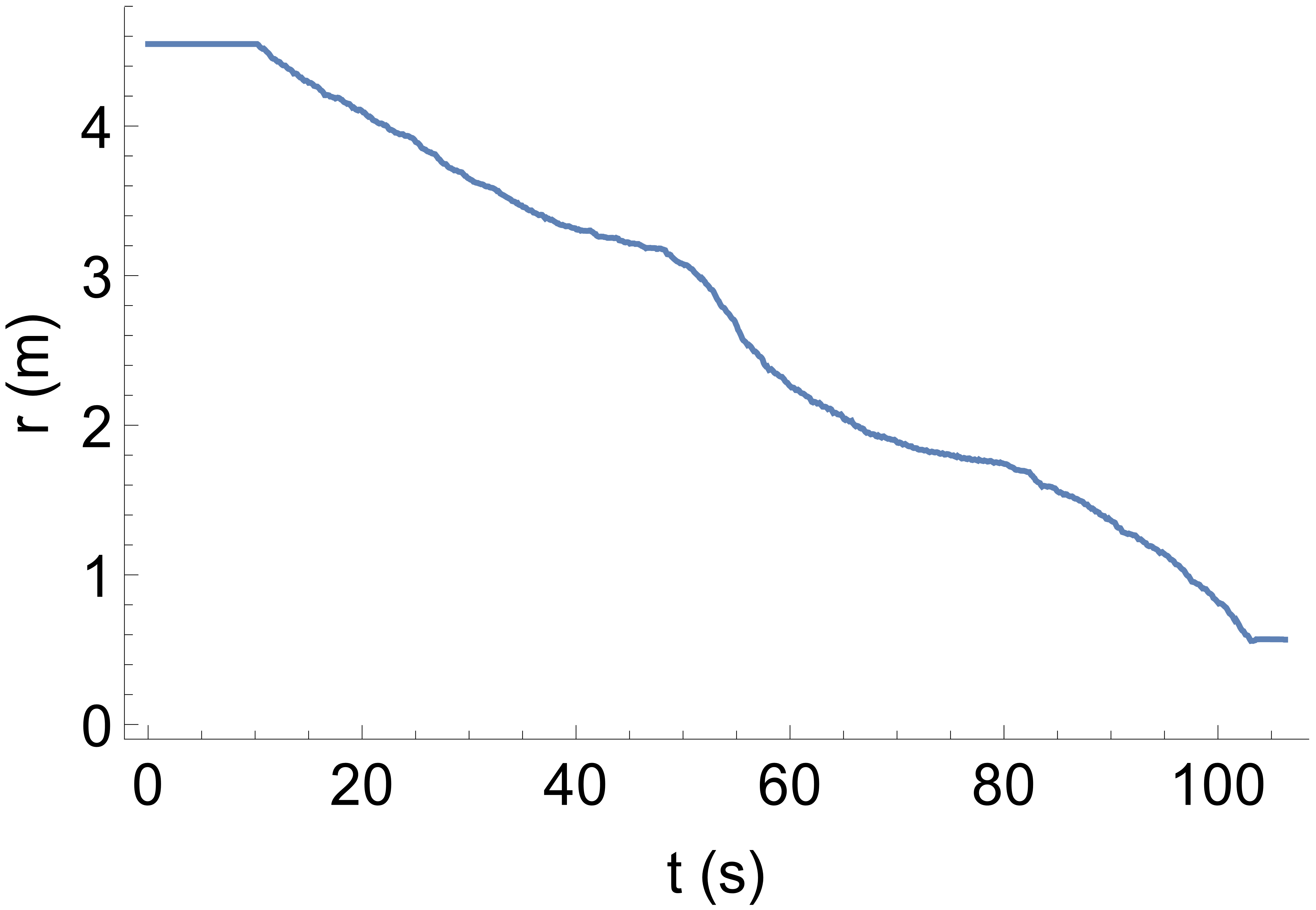}
\includegraphics[width=3in]{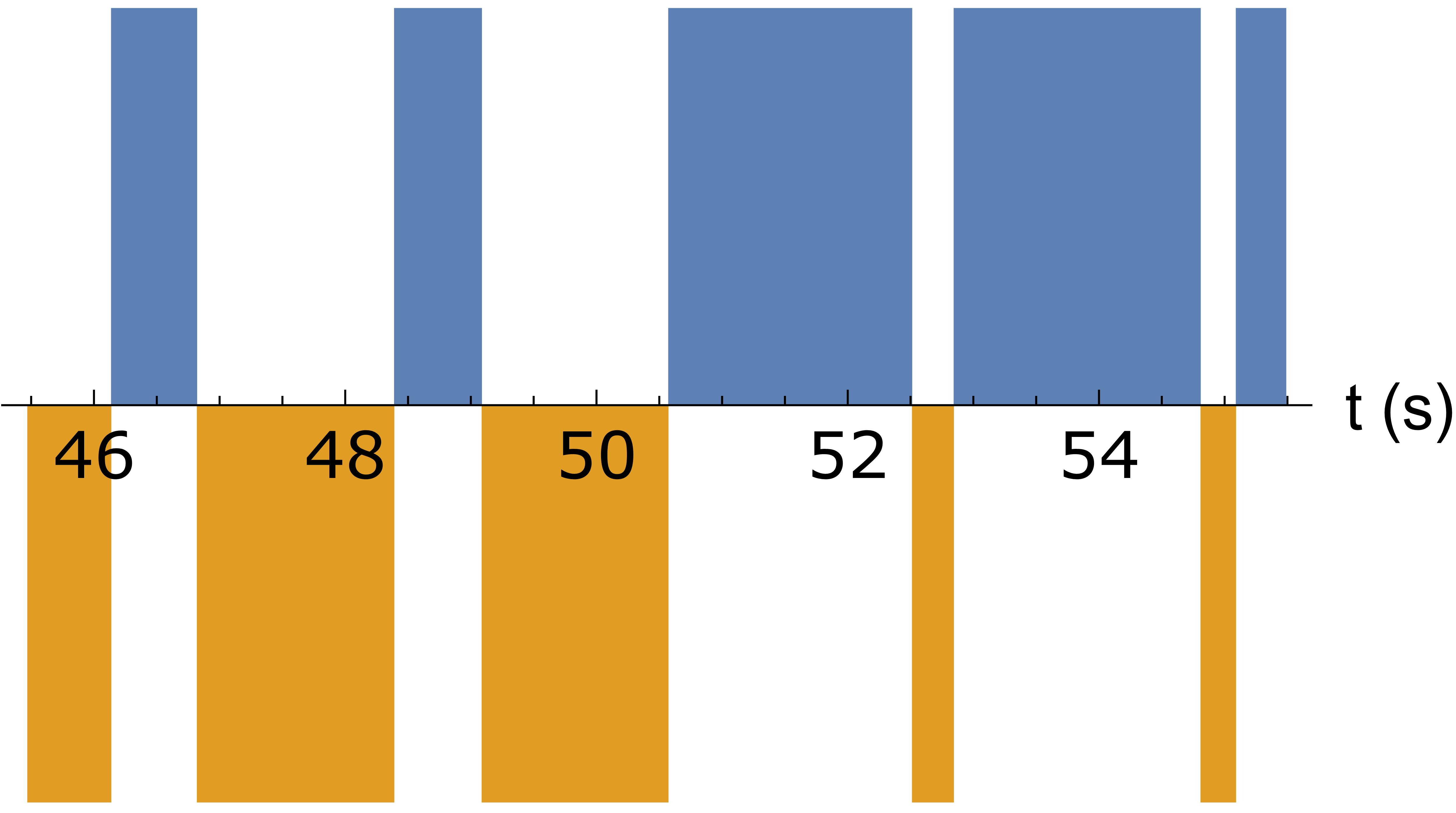}
\captionsetup{justification=raggedright, singlelinecheck=false}
\caption{(a) Distance from the target vs time, calculated for the trajectory of Fig. \ref{fig:trajectory}; \\ 
(b) time series showing intervals when the car is turning left ($+ 1$) and right ($- 1$). This plot 
corresponds to the spike trains of Fig. \ref{fig:spikes_20s}. }
\label{fig:representations}
\end{figure}

Fig. \ref{fig:spikes_20s}b shows, for the same run as in part a), the frequency coming out of the voltage to 
frequency converter, for the L (blue) and R (yellow) circuit. The current injected by the corresponding 
``synapse" into the L / R axon is proportional to this frequency. While the firing rates of the two AAs 
do not need to be perfectly tuned, if, for equal light, one firing rate is larger than the other, this introduces 
a bias in the approach to the light source. In the realization shown in Fig. \ref{fig:trajectory}, the 
right AA had a faster firing rate, and a right turn bias is visible in the trajectory. We come back to 
the question of how much difference in firing rates can be tolerated in the Discussion. \\ 
In Fig. \ref{fig:spikes_all} we report the whole time series for the two AAs corresponding to 
the run of the movie, and in Fig. \ref{fig:representations} two different representations, 
among the many possible ones, of the ``behavior" of the car. \\

\noindent {\bf Materials and Methods.} In this section, we describe in detail 
the EE aspetcs of the demo, and summarize the construction and operation of the AA; 
the latter has been described before \cite{Amila2016, Hector2017}. \\ 

\noindent {\it Molecular biology.}
The KvAP gene in vector pQE60 is expressed in {\it E. coli} strain XL1-Blue competent cells (Aligent) 
and reconstituted in DPhPC vesicles. Vesicles are fused to the supported bilayer to introduce 
the channels in the AA. Protein expression, purification, and reconstitution protocols are described 
at length in previous publications \cite{Hector2017, Amila2016, Amila2013}. \\

\begin{figure}
\centering
\includegraphics[width=3in]{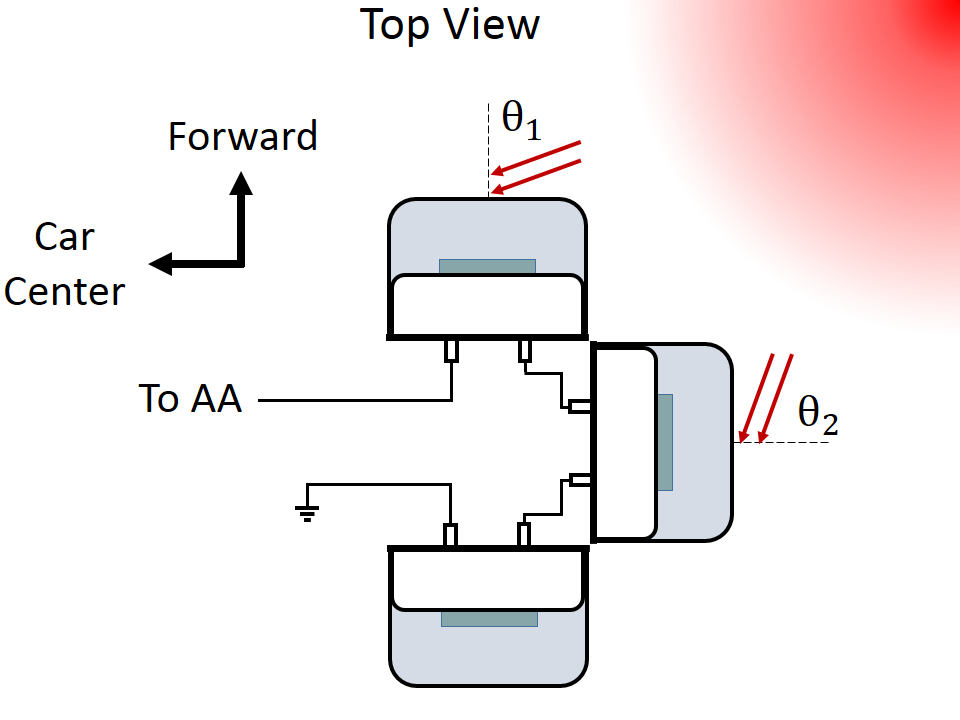}
\captionsetup{justification=raggedright, singlelinecheck=false}
\caption{Top-view schematic of the photodiode arrangement for right-turns. }
\label{fig:car_eyes}
\end{figure}

\noindent {\it The ``eyes" of the car.}
For each eye, three Burr-Brown OPT 301 photodiodes are positioned at 90 degrees relative to each other. 
This arrangement is shown in Figure \ref{fig:car_eyes} for right eye. One photodiode faces the direction of 
forward motion, another is oriented 180 degrees relative to the first in the backward direction, and the third 
is perpendicular to the other two, facing outward. The sum of their outputs is the turn input 
to one artificial axon. 
With this arrangement, if the light source in Figure \ref{fig:car_eyes} was to the left of the car, the right 
photodiodes would output no turning voltage and the car would turn left toward the light source. 

\noindent The voltage output of one photodiode depends on the angle of incidence of the light falling on it. 
For small angles $\theta$ (counted from normal incidence), the voltage drops as $cos(\theta)$, 
and for larger angles the voltage drops 
sigmoidally to zero at $\theta \sim 60$ degrees. As an example, and referring to Fig. \ref{fig:car_eyes}, 
with the light source in the NE direction, the forward and outward-facing photodiodes contribute input 
to the AA for right turns. The backward-facing photodiode is in the dark and therefore does not contribute 
to the AA input. For each eye, the ground pin of one photodiode is connected to the amplifier output 
of the next. The result is one output voltage from all three photodiodes, equal to the sum of their individual outputs. The summed output is converted to a frequency and sent wirelessly to the corresponding 
Artificial Axon. 

\noindent Although it is not necessary, for these navigation experiments we chose the photodiode circuitry 
such that even far away from the light source but at normal incidence, the photodiode output saturates. Specifically, we chose the transimpedance resistance so that the photodiode saturates at 5 meters 
from the 625 nm, 100 W LED light source. \\

\noindent {\it Source and terrain.} 
Heating and background in the high-wattage LED are minimized with an active heat sink and 
brightness shields, respectively. Silicone thermal grease is applied to the interface between the LED and 
a heat sink for improved thermal conduction. Background refers to reflection of light from the floor and walls that adds a constant to the photodiode voltage output and is roughly independent of car orientation. 
The CLVC competes with this background and membrane leaks to keep the axon at the resting potential, 
and struggles when leaks alone are large. A brightness shield above the LED lowers background from reflection off the ceiling and walls, and a shield below the LED reduces background from floor reflection. 
Walls and reflective surfaces are covered with black tarps for background reduction. In total, shielding reduces the background to less than $5 \%$ of the photodiode saturation voltage. \\ 

\noindent {\it Wireless communication.} 
The Burr-Brown VFC32 converts the photodiode voltage output to a digital pulse input for the wireless transmitter. An external resistor value is chosen to make the VFC frequency linearly proportional to the photodiode output, and an external capacitor value is chosen to cap the frequency maximum at 4 kHz. Two separate radio wave transmitters are used for left/right turning, 433 and 315 MHz, one transmitter for each direction. Encoder/decoder components typically paired to transmitters/receivers are excluded here because there is no interference between transmitters, and there is no interference from other devices in the lab where the navigation demonstration is performed. Furthermore, these simple receivers struggle with transmitted data frequencies above 17 kHz, but encoder oscillator frequencies must be larger than this to transmit VFC input in the kHz range. 

\noindent Unlike voltage-to-frequency, the frequency-to-voltage conversion is done by software. 
While the VFC32 can convert frequency to voltage, the wireless receiver outputs a frequency duty cycle 
that is different from the VFC. To correct for the mismatch by hardware is unnecessarily complicated. 
Therefore, the frequency to voltage conversion is instead done by computer using LabVIEW. 
The program converts frequency to voltage, which is then fed to the input of the current clamp which 
forms the ``synapse" injecting into the AA. The command voltage to the clamp is thus 
$V_{cc} = \alpha \, f$ where $f$ is the receiver frequency and $\alpha$ is a proportionality constant. 
There are two independent circuits for the right and left axons. The constant $\alpha$ is chosen so that 
it matches the electrophysiology characteristics of the corresponding axon (mainly dependent on 
leak current, Nernst potential, number of channels) and is therefore different between axons. 
For the run of the movie, we had $\alpha_L = 2.1\, mV/kHz$ for the left axon and $\alpha_R = 1.8\, mV/kHz$ 
for the right axon. The receiver output becomes noisy at frequencies near zero, so an added filter is written into LabVIEW to filter out this frequency noise. LabVIEW interprets this noise as large frequencies as high as 15kHz. The filter removes frequencies higher than 3900Hz, before the voltage conversion.  \\

\begin{figure}
\centering
\includegraphics[width=3in]{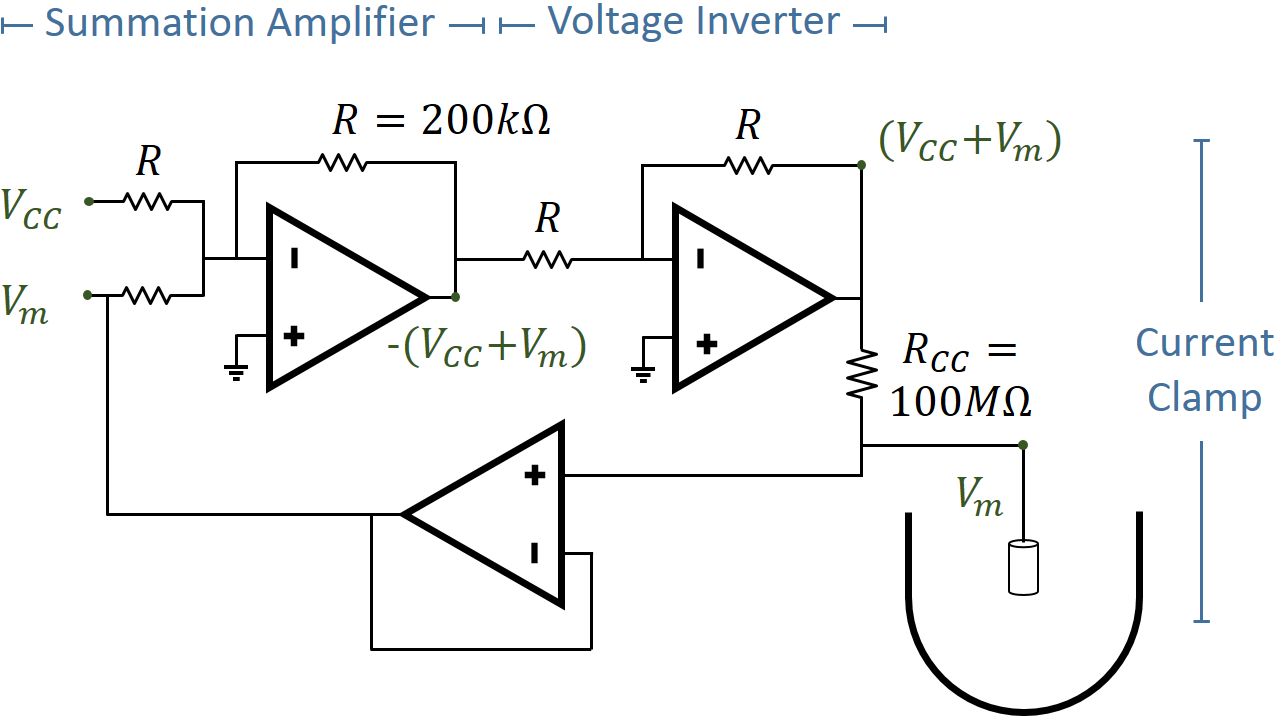}
\captionsetup{justification=raggedright, singlelinecheck=false}
\caption{Schematic of the current clamp. The green dots represent the positions of the stated voltage values, 
also in green. The current clamp has three components: a summation amplifer circuit, a voltage inverter, 
and a high-impedance voltage follower. The potential difference between the voltage inverter and follower 
determines the current injected into the AA as $I_{CC} = V_{CC} / R_{CC}$. }
\label{fig:CC}
\end{figure}

\noindent {\it Current clamp.} 
A schematic of the current clamp is shown in Figure \ref{fig:CC}. All op-amps used are the low-noise FET 
precision op-amp AD795. The summation amplifier adds the command voltage $V_{CC}$ to 
the membrane voltage $V_m$ as $- (V_{CC} + V_m)$, and the inverter flips the voltage reference so that 
the sum is positive. The potential difference across the $100 \, M \Omega$ current clamp resistor 
$R_{CC}$ is $V_{CC}$, and the current injected is $V_{CC} / R_{CC}$. The high-impedance 
voltage follower measures $V_m$ for feedback to the summation amplifier. \\

\noindent {\it CLVC protocol.} 
The CLVC protocol during firing ($V_C, Fig. \ref{fig:action_pot}$) is incidental to the particular inactivation 
dynamics of the KvAP. Channels will completely inactivate if the membrane potential is not pulled down 
to a large negative value after firing. The channel recovery rate from inactivation has a sigmoidal 
dependence on the membrane voltage, with the turning point at about $- 100 \, mV$. Pulling the membrane 
voltage down to $\sim - 100 \, mV$ is typically sufficient to maintain firing. \\ 
For the run of the movie, the following settings were used. For the left axon: 
when the membrane voltage reaches the trigger value $V_T = 4.5 \, mV$, the command voltage 
to the CLVC changes from $V_C (1) = - 127 \, mV$ to $V_C (2) = - 455 \, mV$ 
for $t_T = 1.3 \, s$, pulling the membrane voltage to a large negative value. 
For the right axon: $V_T = 0 \, mV$, $V_C (1) = - 145 \, mV$, $V_C (2) = - 364 \, mV$, 
$t_T = 1.0 \, s$. \\ 
The clamp value $V_C (2)$ is chosen with a big safety margin to address the fact that sometimes 
the leak conductance of the AA changes in the course of a run. For example, you can see in 
Fig. \ref{fig:spikes_all}a, looking at the negative swings of the spikes, that the envelope of the spikes 
is roughly constant (at $\sim - 280 \, mV$) for $0 < t < 80 \, s$, then increases for $80 < t < 100 \, s$, 
then stabilizes again (at $\sim - 120 \, mV$) for $t > 100 \, s$. This increase is caused by an increase 
in leak conductance of the axon, from $\sim (83 \, G \Omega)^{-1}$ to $\sim (2.4 \, G \Omega)^{-1}$, 
approximately. However, even with the increased leak, the same CLVC protocol is able to pull the resting 
voltage down below $- 100 \, mV$, allowing the channels to recover from inactivation and so be able 
to fire repeatedly. Similarly, you see that in the right axon (Fig. \ref{fig:spikes_all}b) the leak conductance 
increases and then decreases again for $20 < t < 50 \, s$. The origin of these slow fluctuations in leak 
conductance is presumably that the interface between lipid bilayer and solid support is not as stable as 
one would wish, in the present system. Similar fluctuations in leak conductance are observed even 
in the absence of channels, so this is a membrane phenomenon. In our present system, a membrane 
with channels lasts typically $\sim 10$ min before it breaks; exceptionally we have lifetimes of $\sim 1$ hour. 
Without channels, a membrane lasts typically 1 hr. Thus we will need to significantly improve the stability 
of the system if we want to scale it up even modestly. \\

\noindent {\it Electrophysiology parameters.} 
For the run of the movie, the axon parameters were set / measured as follows. 
Left axon: at maximum photodiode output, the current clamp injects $I_{CC}^{max} = 74 \, pA$ 
into the axon; the number of open channels at peak voltage is approximately $N = 380$; 
the membrane capacitance is $C = 190 \, pF$. Right axon: $I_{CC}^{max} = 64 \, pA$, 
$N = 720$, and $C = 185 \, pF$. \\ 

\noindent {\it The vehicle.} 
From the manufacturer (GPTOYS), left/right movement on the model S911 car's remote is controlled 
by a $5 \, k \Omega$ 
potentiometer configured as a voltage divider. In our system, we removed the potentiometer 
and connected the car remote to a National Instruments NI USB-6008 data acquisition device (DAQ). 
Analog turn signals are given to the car remote by LabVIEW through the DAQ, in place of the potentiometer. 
The negative terminal of the remote's battery is connected to the DAQ's ground channel, and turn voltages 
are supplied by the DAQ to the remote's ``signal" pin in the voltage divider circuit. \\
\noindent The smallest-radius left turn corresponds to a 0 V signal with respect to ground. The smallest-radius 
right turn corresponds to 3 V, and forward directed wheel orientation corresponds to 1.5 V. Signal values 
between 0 V and 3 V correspond to larger turn radii that decrease linearly as the signal moves away 
from 1.5 V in either direction. When an axon's membrane potential exceeds the set trigger voltage, LabVIEW sends an analog voltage signal to the car remote to make a smallest-radius turn in the direction of 
the axon that fired. The analog turn signal persists until LabVIEW detects a voltage signal 
(from the other axon) to turn in the opposite direction. \\
\noindent From the manufacturer, forward motion of the car is also controlled by a $5 \, k \Omega$ 
potentiometer configured as a voltage divider in the car remote. The stop position corresponds to 
a 3 V signal, and maximum speed corresponds to a 0 V signal. In the remote's circuit protocols 
for forward motion, the applied signal must be at 3 V when the remote is turned on. The car begins to move 
at 150 mV below 3V, i.e. 2.85 V. For compatibility with the slow ($\sim 1 \, Hz$) firing rate of our axons, 
we had to slow down the car. We introduced two modifications. 
First, a LabVIEW function generator supplies square pulses with amplitude $\sim 150 \, mV$ 
and period $500 \, ms$ for forward motion, with an offset chosen so that the maximum voltage 
is $3 \, V$. Second, four 50W resistors are connected in parallel to the car's motor to reduce 
the motor's current. This is a high current RC motor, so the power resistors are necessary. 
These modifications bring the car's speed down to $20 - 30 \, cm/s$. \\

\begin{figure}
\centering
\includegraphics[width=3in]{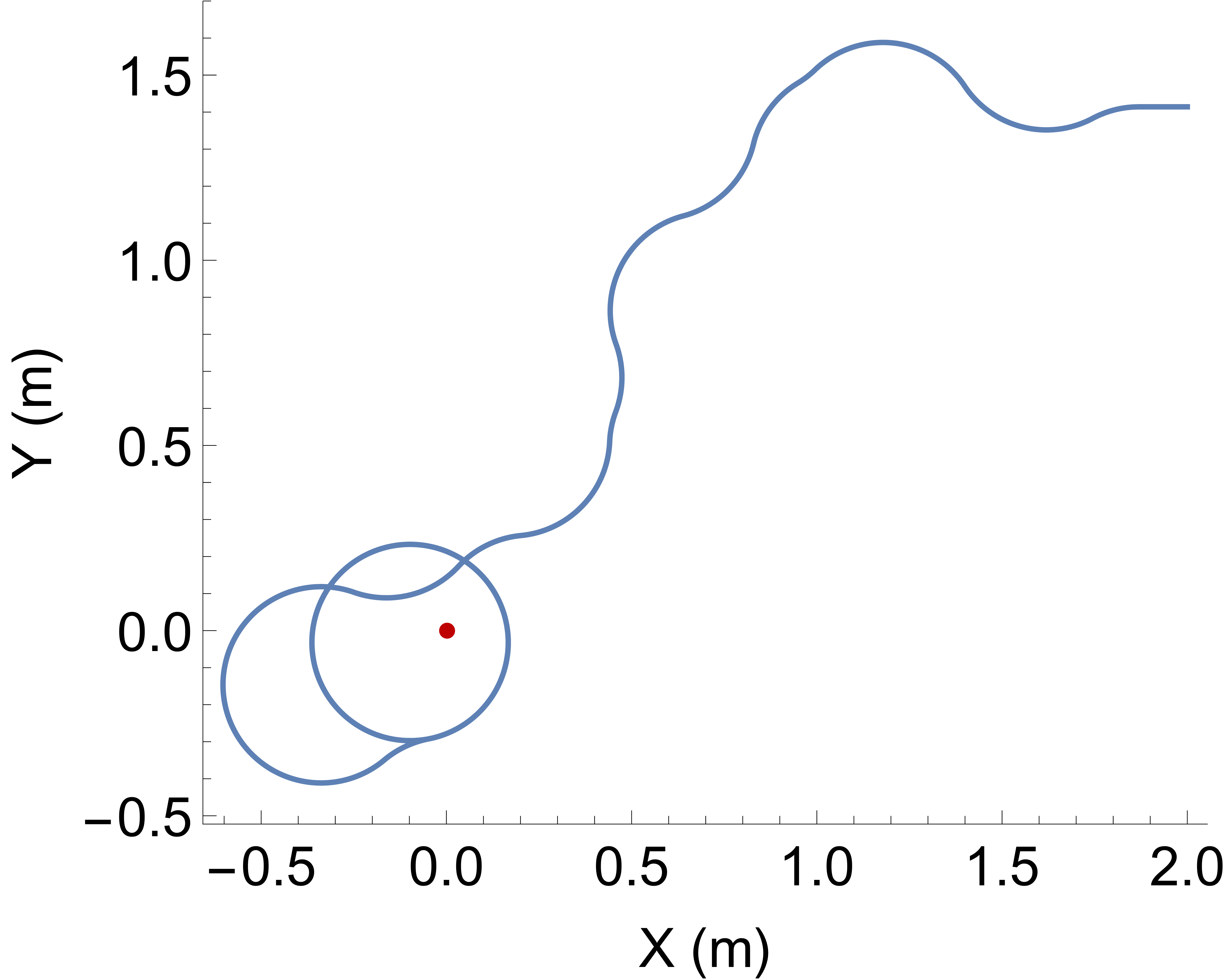}
\captionsetup{justification=raggedright, singlelinecheck=false}
\caption{Car trajectory obtained from a simulation where the right AA has a firing rate 1.9 times higher 
than the left AA, for the same light seen. The car still ``finds" the light source, which is at the origin.}
\label{fig:simulation_trajectory}
\end{figure}

\noindent {\bf Discussion.} Our goal with this demo is to instigate the development 
of ``ionic networks" \cite{Hector2017}. 
We submit that a large network of artificial axons 
connected by tunable synapses would form an interesting neuroscience breadboard.  
One use would be to analyze principles of how the ``microscopics" of action 
potentials may give rise to macroscopic behavior. We note in passing that such a program is  
within the traditional focus of condensed matter physics, which seeks to understand 
``emergent" macroscopic properties starting from the microscopic components and interactions. 
At a higher level of description, the relation between information flow and behavior need not be based 
on complicated rules in order to produce complex behavior. In his delightful book ``Vehicles", 
Valentino Braitenberg explains how simple control mechanisms can lead to surprisingly complex 
behavior \cite{Vehicles}. His very first example in the book is the car with left/right control. However, the 
specific microscopics of action potentials puts constraints on the flow of information and also provides specific 
mechanisms for the interaction of different bits of information. If we believe that the latter process 
is essential for ``thought", we want our test network to be based on nodes which support action potentials. 
Even our simple, non-interacting system is not trivial to analyze, if one gets into a little detail, though 
it is easy enough to simulate. Let us come back to the issue of different firing rates for equal light intensity. 
Fig. \ref{fig:simulation_trajectory} shows the trajectory from a simulation with similar initial conditions 
as the movie (see Materials and Methods for details). In the simulation, the right AA had a firing rate 
1.9 times higher than the left AA, for the same light received at the photodiode. The right turn bias is 
visible in the car's trajectory, but overall the car still finds the light source. The end state is a limit cycle which 
is a circle containing the light source. How much difference in the firing rates can be tolerated depends 
on the other parameters of the system. For example, with the firing rate $\nu$, the car speed $u$, 
the turn radius of the car $r$, the initial distance to the light source $L$, we can form the two dimensionless 
numbers $\chi = u / (\nu r)$ and $\rho = r/L$. Then we can discuss, in this parameter space, 
the basin of attraction of the set of limit cycles which form the desirable end states. 
However, this is already a complicated question to explore analytically, for such a simple dynamical system ! 

Looking to the future, we are far from being able to construct a self-contained ionic network. 
Some of the difficulties seem surmountable with present day engineering, others would 
require new inventions. In the context of this demo, for example, we can see a path 
for substituting some of the electronic components with ionics. The CLVC could be 
dispensed with by adding a second ionic gradient, e.g. of $Na^+$, and corresponding 
voltage gated ion channels. Photodiodes could in principle be replaced by AAs with embedded 
channel rhodopsin. On the other hand, ionics based ``synapses" compatible with the 
$100 \, mV$ scale of ionic action potentials require new inventions. Finally, 3D printing technology 
currently being developed to produce scaffolds for directed neuronal growth \cite{3Dprint_MIT} 
could probably form the basis for scaling up our AA network. \\

\section{Acknowledgements}
\noindent This work was supported by funds from the Dean of Physical Sciences at UCLA. 
We thank Albert Libchaber for pointing us to Valentino Breitenberg's book.

\bibliography{Hector_2}

\end{document}